\documentstyle[12pt]{article}
\def\zid{1\kern-0.36em\llap~1}

\newcommand{\beq}{\begin{equation}}
\newcommand{\ber}{\begin{eqnarray}}
\newcommand{\eeq}{\end{equation}}
\newcommand{\eer}{\end{eqnarray}}

\begin{document}

\begin{titlepage}
\rightline{[SUNY BING 4/17/03 v2] }
 \rightline{ hep-ph/0304198}
\vspace{2mm}
\begin{center}
{\bf \hspace{0.1 cm} ON THE TOP QUARK'S CHIRAL WEAK-MOMENT }\\
\vspace{2mm} Charles A. Nelson\footnote{Electronic address:
cnelson @ binghamton.edu  } \\ {\it Department of Physics, State
University of New York at Binghamton\\ Binghamton, N.Y.
13902}\\[2mm]
\end{center}


\begin{abstract}

For the observed  $t \rightarrow W^{+} b $ decay, an
intensity-ratio equivalence-theorem for two Lorentz-invariant
couplings is shown to be related to symmetries of
tWb-transformations. Explicit tWb-transformations, $A_{+}=M $
$A_{SM}$, $P $ $A_{SM}$, $B $ $A_{SM}$ relate the four standard
model's helicity amplitudes, $A_{SM}\left( \lambda _{W^{+} }
,\lambda _b \right)$, and the amplitudes $A_{+}\left( \lambda
_{W^{+} } ,\lambda _b \right)$ in the case of an additional $t_R
\rightarrow b_L$ weak-moment of relative strength $\Lambda_{+}
=E_W /2 \sim 53 GeV$. Two ``commutator plus anti-commutator"
symmetry algebras are generated from $ M, P, B$. These
transformations enable a simple and uniform characterization of
the values of $\Lambda_{+}$, $m_W/m_t$, and $m_b/m_t$.

\end{abstract}

\end{titlepage}

\section{Introduction:}

In this paper, for the observed $t \rightarrow W^{+} b $ decay
[1], an intensity-ratio equivalence-theorem [2] for two
Lorentz-invariant couplings is shown to be related to symmetries
of tWb-transformations, $A_{+}=M $ $A_{SM}$, $P$ $A_{SM}$, $B $
$A_{SM}$, where $ M, P, B$ are explicit $4$x$4$ matrices. These
\newline tWb-transformations relate the standard model's helicity
amplitudes, $A_{SM}\left( \lambda _{W^{+} } ,\lambda _b \right)$,
and the amplitudes $A_{+}\left( \lambda _{W^{+} } ,\lambda _b
\right)$ in the case of an additional $t_R \rightarrow b_L$
weak-moment of relative strength $\Lambda_{+} =E_W /2 \sim 53
GeV$.  Versus the standard model's pure $(V-A)$ coupling, the
additional tensorial coupling can be physically interpreted as
arising due to a large chiral weak-transition moment for the top
quark. $\Lambda_{+}$ is defined by (1) below and the $(+)$
amplitudes' complete coupling is (2). $E_W$ is the energy of the
final W-boson in the decaying top-quark rest frame. The subscripts
$R$ and $L$ respectively denote right and left chirality of the
coupling, that is $( 1 \pm \gamma_5 ) $. $\lambda _{W^{+} }$, $
\lambda _b $ are the helicities of the the emitted W-boson and
b-quark in the top-quark rest frame.   The Jacob-Wick
phase-convention [3] is used in specifying the phases of the
helicity amplitudes and so of these transformations.

Due to rotational invariance, there are four independent $A\left(
\lambda _{W^{+} } ,\lambda _b \right)$ amplitudes for the most
general Lorentz coupling [4,5]. Stage-two spin-correlation
functions were derived and studied as a basis for complete
measurements of the helicity parameters for $t \rightarrow W^{+} b
$ decay as tests with respect to the most general Lorentz
coupling.  Such tests are possible at the Tevatron [1], at the LHC
[6], and at a NLC [7]. In this paper, a subset of the most general
Lorentz coupling is considered in which the subscript ``$i$"
identifies the amplitude's associated coupling: ``$i=$ SM" for the
pure $(V-A)$ coupling, ``$i=(f_M + f_E)$" for only the additional
$t_R \rightarrow b_L$ tensorial coupling, and ``$i=(+)$" for
$(V-A) + (f_M + f_E)$ with a top-quark chiral weak-transition
moment of relative strength $\Lambda_{+} =E_W /2$ versus $g_L$.
The Lorentz coupling involving both the SM's $(V-A)$ coupling and
an additional $t_R \rightarrow b_L $ weak-moment coupling of
arbitrary relative strength $\Lambda_{+}$ is $ W_\mu ^{*} J_{\bar
b t}^\mu = W_\mu ^{*}\bar u_{b}\left( p\right) \Gamma ^\mu u_t
\left( k\right) $ where $k_t =q_W +p_b $, and
\begin{equation}
\frac{1}{2} \Gamma ^\mu =g_L\gamma ^\mu P_L + \frac{g_{f_M + f_E}
} {2\Lambda _{+} }\iota \sigma ^{\mu \nu } (k-p)_\nu P_R
\end{equation}
where $P_{L,R} = \frac{1}{2} ( 1 \mp \gamma_5 ) $.  In $g_L =
g_{f_M + f_E} = 1$ units, when $\Lambda_{+} = E_W/2$ which
corresponds to the (+) amplitudes, the complete $t \rightarrow b$
coupling is very simple
\begin{equation}
\gamma ^\mu P_L + \iota \sigma ^{\mu \nu } v_\nu P_R
 =P_R \left( \gamma
^\mu + \iota \sigma ^{\mu \nu } v_\nu \right)
\end{equation}
where $v_{\nu}$ is the W-boson's relativistic four-velocity.

In the  $t$ rest frame, the helicity-amplitude matrix element for
$t \rightarrow W^{+} b$ is \newline $ \langle \theta _1^t ,\phi
_1^t ,\lambda _{W^{+} } ,\lambda _b |\frac 12,\lambda _1\rangle =$
 $ D_{\lambda _1,\mu }^{(1/2)*}(\phi _1^t ,\theta _1^t ,0)A_{i}
\left( \lambda _{W^{+} } ,\lambda _b \right) $ where $\mu =\lambda
_{W^{+} } -\lambda _b $ in terms of the $W^+$ and $b$-quark
helicities.  The asterisk denotes complex conjugation, the final
$W^{+}$ momentum is in the $\theta _1^t ,\phi _1^t$ direction, and
$\lambda_1$ gives the $t$-quark's spin component quantized along
the $z$ axis. $\lambda_1$ is also the helicity of the $t$-quark if
one has boosted, along the ``$-z$" direction, back to the $t$ rest
frame from the $(t \bar{t})_{cm}$ frame.  It is this boost which
defines the $z$ axis in the $t$-quark rest frame for angular
analysis [4].  Explicit expressions for the helicity amplitudes
associated with each ``$i$" coupling are listed in Sec. 2. We
denote by $\Gamma$ the partial-width for the $t \rightarrow W^{+}
b $ decay channel and by $\Gamma_{L,T}$ the partial-width's for
the sub-channels in which the $W^{+}$ is respectively
longitudinally, transversely polarized; $\Gamma = \Gamma_L
+\Gamma_T $.  Similarly, ${\Gamma_L}|_{\lambda_b = - \frac{1}{2}}
$ denotes the partial-width for the W-longitudinal sub-channel
with b-quark helicity $\lambda_b = - \frac{1}{2}$.

The intensity-ratio equivalence-theorem states, ``As consequence
of Lorentz-invariance,  for the $t \rightarrow W^{+} b $ decay
channel each of the four ratios ${\Gamma_L}|_{\lambda_b = -
\frac{1}{2}} / {\Gamma}$, ${\Gamma_T}|_{\lambda_b = - \frac{1}{2}}
/ {\Gamma}$, ${\Gamma_L}|_{\lambda_b =  \frac{1}{2}} / {\Gamma}$,
${\Gamma_T}|_{\lambda_b =  \frac{1}{2}} / {\Gamma}$, is identical
for the pure $(V-A)$ coupling and for the $(V-A) + (f_M + f_E)$
coupling with $\Lambda_{+} =E_W /2$, and their respective
partial-widths are related by $\Gamma_{+} = v^2 \Gamma_{SM}$." $v$
is the velocity of the W-boson in the t-quark rest frame. Note
that this theorem does not require specific values of the mass
ratios $y \equiv m_W/m_t$, and $x \equiv m_b/m_t$, but that the
relative strength of the chiral weak-transition moment for the top
quark has been fixed versus $g_L$.

The three tWb-transformations, $A_{+}=M $ $A_{SM}$, $P $ $A_{SM}$,
$B $ $A_{SM}$, are related to this theorem.  The $M$
transformation implies the theorem, but as explained below, $M$
also implies the sign and ratio differences of the (ii) and (iii)
type amplitude ratio-relations which distinguish the (SM) and (+)
couplings. The $P$ and $B$ transformations more completely exhibit
the underlying symmetry relating these two Lorentz-invariant
couplings. In particular, these three $4$x$4$ matrices lead to two
``commutator plus anti-commutator" symmetry algebras, and together
enable a simple and uniform characterization of the values of
$\Lambda_{+}$, $y \equiv m_W/m_t$, and $x \equiv m_b/m_t$.  In
Sec. 2, it is shown how these three tWb-transformations
successively arise from consideration of different types of
``helicity amplitude relations" for $t \rightarrow W^{+} b $
decay: The type (i) are ratio-relations which hold separately for
the two cases, ``$i=(SM)$, $(+)$". The type (ii) are
ratio-relations which relate the amplitudes in the two cases.  By
the type (iii) ratio-relations, the tWb-transformation $A_{+}=M$
$A_{SM}$ where $M=v$ $diag(1,-1,-1,1)$ characterizes the mass
scale $\Lambda_{+} = E_W/2 $. Similarly, the amplitude condition
(iv)
\begin{equation}
A_{+} (0,-1/2) = a A_{SM} (-1,-1/2),
 \end{equation}
with $a= 1 + O(v \neq y \sqrt{2}, x)$, determines the scale of the
tWb-transformation matrix $P$ and determines the value of the mass
ratio $y \equiv m_W/m_t$. $O(v \neq y \sqrt{2}, x)$ denotes small
corrections, see below. The amplitude condition (v)
\begin{equation}
A_{+} (0,-1/2) = - b A_{SM} (1-1/2),
\end{equation}
with $ b = v^{-8} $, determines the scale of $B$ and determines
the value of $ x= m_b/m_t$.  In Sec. 3, two symmetry algebras are
obtained which involve the $M$, $P$, and $B$ transformation
matrices. Sec. 4 contains some remarks.

\section{Helicity amplitude relations:}

In the Jacob-Wick phase convention, the helicity amplitudes for
the most general Lorentz coupling are given in [4]. In $g_L =
g_{f_M + f_E} = 1$ units and suppressing a common overall factor
of $\sqrt{m_t \left( E_b +q_W \right) }$, for only the $(V-A)$
coupling the associated helicity amplitudes are:
\begin{eqnarray*}
 A_{SM} \left(
0,-\frac 12\right) & = & \frac{1 }{y } \; \frac{E_W +q_W }{m_t }
\\
 A_{SM} \left(
-1,-\frac 12\right) & = & \sqrt{2}  \\
 A_{SM} \left( 0,\frac 12\right)
& = & -  \frac{1 }{y }  \frac{E_W -q_W }{m_t } \left( \frac
{m_b}{m_t-E_W +  q_W}  \right) \\
 A_{SM} \left( 1,\frac 12\right) & = & -
\sqrt{2} \left( \frac {m_b}{m_t-E_W +  q_W}  \right)
\end{eqnarray*}
For only the $(f_M + f_E)$ coupling, i.e. only the additional $t_R
\rightarrow b_L $ tensorial coupling:
\begin{eqnarray*}
 A_{f_M + f_E} \left(
0,-\frac 12\right) & = &  - ( \frac{m_t }{2\Lambda_+ }) \;  y  \\
 A_{f_M + f_E} \left(
-1,-\frac 12\right) & = & - ( \frac{m_t }{2\Lambda_+ }) \sqrt{2}
\; \frac{E_W +q_W }{m_t } \\
 A_{f_M + f_E} \left( 0,\frac 12\right)
& = & ( \frac{m_t }{2\Lambda_+ }) y \left( \frac {m_b}{m_t-E_W +
q_W}  \right) \\
 A_{f_M + f_E} \left( 1,\frac 12\right) & = & ( \frac{m_t }{2\Lambda_+ })
\sqrt{2} \;  \frac{E_W - q_W }{m_t } \left( \frac {m_b}{m_t-E_W +
 q_W}  \right)
\end{eqnarray*}
From these, the amplitudes for the $(V-A) + (f_M + f_E)$ coupling
of (1) are obtained by \newline $A_{+} (\lambda_W, \lambda_b) =
A_{SM} (\lambda_W, \lambda_b) + A_{f_M + f_E } (\lambda_W,
\lambda_b)$. For $\Lambda_{+} = E_W /2$, the $A_{+} (\lambda_W,
\lambda_b)$ amplitudes [8] corresponding to the complete $t
\rightarrow b$ coupling (2) are
\begin{eqnarray*}
 A_{+} \left(
0,-\frac 12\right) & = & \frac{1 }{y } \; (q/E_W)  \; \frac{E_W
+q_W }{m_t }
\\
 A_{+} \left(
-1,-\frac 12\right) & = & - \sqrt{2} \; (q/E_W) \\
 A_{+} \left( 0,\frac 12\right)
& = & \frac{1 }{y } \; (q/E_W) \;  \frac{E_W -q_W }{m_t } \left(
\frac {m_b}{m_t-E_W +  q_W}  \right) \\
 A_{+} \left( 1,\frac 12\right) & = & -
\sqrt{2} \; (q/E_W) \; \left( \frac {m_b}{m_t-E_W +  q_W}  \right)
\end{eqnarray*}

We now analyze the different types of helicity amplitude relations
involving both the SM's amplitudes and those in the case of the
$(V-A) + (f_M + f_E)$ coupling: The first type of ratio-relations
holds separately for $i=(SM)$, $(+)$ and for all $y = \frac {m_W}
{m_t} , x = \frac {m_b} {m_t} , \Lambda_{+}$ values, (i):
\begin{equation}
\frac{A_{i} (0,1/2) } { A_{i} (-1,-1/2) } = \frac{1}{2}
\frac{A_{i} (1,1/2) } { A_{i} (0,-1/2) }
\end{equation}

The second type of ratio-relations relates the amplitudes in the
two cases and also holds for all $y, x, \Lambda_{+}$ values. The
first two relations have numerators with opposite signs and
denominators with opposite signs, c.f. Table 1; (ii): Two
sign-flip relations
\begin{equation}
\frac{A_{+} (0,1/2) } { A_{+} (-1,-1/2) } = \frac{A_{SM} (0,1/2) }
{ A_{SM} (-1,-1/2) }
\end{equation}
\begin{equation}
\frac{A_{+} (0,1/2) } { A_{+} (-1,-1/2) } = \frac{1}{2}
\frac{A_{SM} (1,1/2) } { A_{SM} (0,-1/2) }
\end{equation}
and two non-sign-flip relations
\begin{equation}
\frac{A_{+} (1,1/2) } { A_{+} (0,-1/2) } = \frac{A_{SM} (1,1/2) }
{ A_{SM} (0,-1/2) }
\end{equation}
\begin{equation}
\frac{A_{+} (1,1/2) } { A_{+} (0,-1/2) } = 2 \frac{A_{SM} (0,1/2)
} { A_{SM} (-1,-1/2) }
\end{equation}
Eqs(7,9), which are not in [2], are essential for obtaining the
$P$ and $B$ tWb-transformations and thereby the symmetry algebras
of Sec. 3 below.

The third type of ratio-relations, holding for all $y, x$ values,
follows by determining the effective mass scale, $\Lambda_{+}$, so
that there is an exact equality for the ratio of left-handed
amplitudes (iii):
\begin{equation}
\frac{A_{+} (0,-1/2) } { A_{+} (-1,-1/2) } = -
 \frac{A_{SM} (0,-1/2) } { A_{SM} (-1,-1/2) },
\end{equation}
Equivalently, $ \Lambda_{+} = \frac{m_t } {4 } [ 1 + (m_W / m_t)^2
- (m_b / m_t)^2] = E_W/2$ follows from each of:
\begin{equation}
\frac{A_{+} (0,-1/2) } { A_{+} (-1,-1/2) } = -
 \frac{1}{2} \frac{A_{SM} (1,1/2) } { A_{SM} (0,1/2) },
\end{equation}
\begin{equation}
\frac{A_{+} (0,1/2) } { A_{+} (1,1/2) } = - \frac{A_{SM} (0,1/2) }
{ A_{SM} (1,1/2) },
\end{equation}
\begin{equation}
\frac{A_{+} (0,1/2) } { A_{+} (1,1/2) } = - \frac{1}{2}
\frac{A_{SM} (-1,-1/2) } { A_{SM} (0,-1/2) },
\end{equation}
From the amplitude expressions given above, the value of this
scale $\Lambda_{+}$ can be characterized by postulating the
existence of a tWb-transformation $A_{+}=M$ $A_{SM}$ where $M=v$
$diag(1,-1,-1,1)$, with $
A_{SM}=[A_{SM}(0,-1/2),A_{SM}(-1,-1/2),A_{SM}(0,1/2),A_{SM}(1,1/2)]$
and analogously for $A_{+}$.

Assuming (iii), the fourth type of relation is the equality (iv):
\begin{equation}
A_{+} (0,-1/2) = a A_{SM} (-1,-1/2),
\end{equation}
where $a= 1 + O(v \neq y \sqrt{2}, x)$.

This is equivalent to the velocity formula $  v = a  y \sqrt{2}
\left( \frac{1} {1- ( E_b - q_W )/m_t} \right)
 = a y \sqrt{2} $ for $ m_b = 0 $.
In [2], for $a=1$ it was shown that (iv) leads to a mass relation
with the solution $y=\frac {m_W} {m_t} = 0.46006$ ($ x=0$). The
present empirical value is $ y = 0.461 \pm 0.014$, where the error
is dominated by the $3 \% $ precision of $m_t$. In [2], for $a=1$
it was also shown that (iv) leads to $\sqrt{2}=v\gamma
(1+v)=v\sqrt{\frac{1+v}{1-v}}$ so $v=0.6506\ldots$ without input
of a specific value for $m_b$. However, by Lorentz invariance $v$
must depend on $m_b$. Accepting (iii), we interpret this to mean
that $a \neq 1$ and in the Appendix obtain the form of the $O(v
\neq y \sqrt{2}, x)$ corrections in $a$ as required by Lorentz
invariance.  The small correction $O(v \neq y \sqrt{2}, x)$
depends on both $x \equiv m_b/m_t$ and the difference $v-y
\sqrt{2}$.

Equivalently, by use of (i)-(iii) relations, (14) can be expressed
postulating the existence of a second tWb-transformation $A_{+}=P
$ $A_{SM}$ where
\begin{equation}
P\equiv v
\left[
\begin{array}{cccc}
0 & a/v & 0 & 0 \\ -v/a & 0 & 0 & 0 \\ 0 & 0 & 0 & -v/2a \\ 0 & 0
& 2a/v & 0
\end{array}
\right]
\end{equation}
The value of the parameter $a$ of (iv) is not fixed by (15).

The above two tWb-transformations do not relate the $\lambda_b= -
\frac{1}{2}$ amplitudes with the $\lambda_b= \frac{1}{2}$
amplitudes. From (i) thru (iv), in terms of a parameter $b$, the
equality (v):
\begin{equation}
A_{+} (0,-1/2) = - b A_{SM} (1,1/2),
\end{equation}
is equivalent to $A_{+}=B $ $A_{SM}$
\begin{equation}
B\equiv
\left[
\begin{array}{cccc}
0 & 0 & 0 & -b \\ 0 & 0 & 2b & 0 \\ 0 & v^{2}/2b & 0 & 0
\\ -v^{2}/b & 0 & 0 & 0
\end{array}
\right]
\end{equation}
The choice of $ b = v^{-8} = 31.152$, gives
\begin{equation} B\equiv v
\left[
\begin{array}{cccc}
0 & 0 & 0 & -v^{-9} \\ 0 & 0 & 2v^{-9} & 0 \\ 0 & v^{9}/2 & 0 & 0
\\ -v^{9} & 0 & 0 & 0
\end{array}
\right]
\end{equation}
and corresponds to the mass relation $ m_{b}
=\frac{m_{t}}{b}\left[ 1-\frac{vy}{\sqrt{2}}\right] =4.407...GeV $
for $m_t = 174.3 GeV$.

\section{Commutator plus anti-commutator symmetry algebras:}

The anti-commuting matrices
\begin{equation}
m\equiv \left[
\begin{array}{cc}
1 & 0 \\ 0 & -1
\end{array}
\right] ,p\equiv \left[
\begin{array}{cc}
0 & -a/v \\ v/a & 0
\end{array}
\right] ,q\equiv \left[
\begin{array}{cc}
0 & a/v \\ v/a & 0
\end{array}
\right]
\end{equation}
satisfy $[m,p]=-2q,[m,q]=-2p,[p,q]=-2m$. \ Similarly, $m$ and
\begin{equation}
r\equiv \left[
\begin{array}{cc}
0 & -v/2a \\ 2a/v & 0
\end{array}
\right] ,s\equiv \left[
\begin{array}{cc}
0 & v/2a \\ 2a/v & 0
\end{array}
\right]
\end{equation}
are anti-commuting and satisfy $%
[m,r]=-2s,[m,s]=-2r,[r,s]=-2m$. Note $m^2=q^2=s^2=1$,
$p^2=r^2=-1$, and that $a$ is arbitrary.  Consequently, if one
does not distinguish the $(+)$ versus SM indices, respectively of
the rows and columns, the tWb-transformation matrices have some
simple properties:

The anticommuting 4x4 matrices
\begin{equation}
M\equiv v\left[
\begin{array}{cc}
m & 0 \\ 0 & -m
\end{array}
\right] ,P\equiv v\left[
\begin{array}{cc}
-p & 0 \\ 0 & r
\end{array}
\right] ,Q\equiv v\left[
\begin{array}{cc}
q & 0 \\ 0 & s
\end{array}
\right]
\end{equation}
satisfy the closed algebra
$[\overline{M},\overline{P}]=2\overline{Q},[\overline{M},\overline{Q}]
=2\overline{P},[\overline{P},\overline{Q}]=2\overline{M}$. The bar
denotes removal of the overall ``$v$" factor, $M= v \overline{M},
...$.  Note that $Q$ is not a tWb-transformation.

Including the B matrix with $b$ arbitrary, the algebra closes with
3 additional matrices
\begin{equation}
\overline{B}\equiv \left[
\begin{array}{cc}
0 & d \\ f & 0
\end{array}
\right] ,\overline{C}\equiv \left[
\begin{array}{cc}
0 & e \\ g & 0
\end{array}
\right]
\end{equation}

\begin{equation}
\overline{G}\equiv \left[
\begin{array}{cc}
0 & h \\ k & 0
\end{array}
\right] ,\overline{H}\equiv \left[
\begin{array}{cc}
0 & j \\ l & 0
\end{array}
\right]
\end{equation}
where
\begin{equation}
d\equiv \left[
\begin{array}{cc}
0 & -b/v \\ 2b/v & 0
\end{array}
\right] ,e\equiv \left[
\begin{array}{cc}
0 & b/v \\ 2b/v & 0
\end{array}
\right] ,f\equiv \left[
\begin{array}{cc}
0 & v/2b \\ -v/b & 0
\end{array}
\right] ,g\equiv \left[
\begin{array}{cc}
0 & v/2b \\ v/b & 0
\end{array}
\right]
\end{equation}

\begin{equation}
h\equiv \left[
\begin{array}{cc}
-2ab/v^{2} & 0 \\ 0 & b/a
\end{array}
\right] ,j\equiv \left[
\begin{array}{cc}
2ab/v^{2} & 0 \\ 0 & b/a
\end{array}
\right] ,
 \newline
k\equiv \left[
\begin{array}{cc}
1/2v^{2}ab & 0 \\ 0 & -a/b
\end{array}
\right] ,l\equiv \left[
\begin{array}{cc}
1/2v^{2}ab & 0 \\ 0 & a/b
\end{array}
\right]
\end{equation}
The squares of the $2$x$2$ matrices (24-25) do depend on $a$, $b$,
and $v$.

The associated closed algebra is: $[\overline{M},\overline{B}]=0,\{\overline{%
M},\overline{B}\}=-2\overline{C};[\overline{B},\overline{C}]=0,\{\overline{B}%
,\overline{C}\}=-2\overline{M};$ \newline
$[\overline{M},\overline{C}]=0,\{\overline{M},%
\overline{C}\}=-2\overline{B};$ and
$[\overline{P},\overline{B}]=2\overline{H}
,\{\overline{P},\overline{B}\}=0;[%
\overline{H},\overline{P}]=2\overline{B},\{\overline{H},\overline{P}\}=0;$
\newline
$[%
\overline{H},\overline{B}]=2\overline{P},\{\overline{H},\overline{B}\}=0$
. Similarly,  $[\overline{P},\overline{C}]=0,\{\overline{P},\overline{%
C}\}=-2\overline{G};[\overline{M},\overline{H}]=-2\overline{G},$
\newline
$ \{\overline{M}%
,\overline{H}\}=0;[\overline{H},\overline{C}]=0,\{\overline{H},\overline{C}%
\}=2\overline{Q};$ and
$[\overline{M},\overline{G}]=-2\overline{H},\{%
\overline{M},\overline{G}\}=0;[\overline{P},\overline{G}]=0,$
\newline
$\{\overline{P},\overline{G}\}=2\overline{C};[\overline{G},\overline{B}]=-2%
\overline{Q},\{\overline{G},\overline{B}\}=0;$ and  $[\overline{G},\overline{C%
}]=0,\{\overline{G},\overline{C}\}=-2\overline{P};$ \newline $[\overline{G},\overline{H}%
]=2\overline{M},\{\overline{G},\overline{H}\}=0.$ The part involving  $%
\overline{Q}$ \  is
$[\overline{G},\overline{Q}]=2\overline{B},\{\overline{G},\overline{Q}\}=0;[%
\overline{B},\overline{Q}]=2\overline{G},$ \newline
 $ \{\overline{B},\overline{Q}\}=0;
 [\overline{C},\overline{Q}]=0,
 \{\overline{C},\overline{Q}\}=-2\overline{H};$ $%
[\overline{H},\overline{Q}]=0,\{\overline{H},\overline{Q}\}=
2\overline{C}$.

\bigskip

\ This has generated an additional tWb-transformation $G\equiv v\overline{G}$%
; but $C\equiv v\overline{C}$ and $H\equiv v\overline{H}$ are not
tWb-transformations. \

\bigskip

Up to the insertion of an overall $\iota =\sqrt{-1}$, each of
these 4x4 barred matrices is a resolution of unity, i.e.
$\overline{P}^{-1}=-\overline{P}$,
$\overline{G}^{-1}=-\overline{G}$, but
$\overline{Q}^{-1}=\overline{Q}$,
$\overline{B}^{-1}=\overline{B},...$ .

\section{Remarks:}

\indent {\bf (1)  Summary:} The elements of the three
logically-successive tWb-transformations are constrained by the
helicity amplitude ratio-relations (i) and (ii). Thereby, the type
(iii) ratio-relation fixes $\Lambda_{+} = E_W/2$ and the overall
scale of the tWb-transformation matrix $M$. The amplitude
condition (iv), $A_{+} (0,-1/2) = a A_{SM} (-1,-1/2)$ with $a= 1 +
O(v \neq y \sqrt{2}, x)$, and the amplitude condition (v), $A_{+}
(0,-1/2) = - b A_{SM} (1-1/2)$ with $ b = v^{-8} $, determine
respectively the scale of the tWb-transformation matrices $P$ and
$B$ and characterize the values of $m_W/m_t$ and $m_b/m_t$. The
overall scale can be set here by $ m_t $ or $ {m_W}$. From an
empirical ``bottom-up" perspective of further ``unification", $
m_W$ is more appropriate to use to set the scale since its value
is fixed in the SM.

{\bf (2) Symmetries and Second Class Currents:} The additional
$t_R \rightarrow b_L$ weak-moment coupling violates the
conventional gauge invariance transformations of the SM and
traditionally in electroweak studies such anomalous couplings have
been best considered as ``induced" or ``effective". Nevertheless,
in special ``new physics" circumstances such a simple tensorial
coupling as (2) might turn out to be fundamental. The ``tensorial
coupling" is of course a fundamental structure if considered from
gravitation viewpoints. However, are the new symmetries associated
with the symmetry algebras of Sec. 3 sufficient to overcome the
known difficulties [9] in constructing a renormalizable, unitary
quantum field theory involving second class currents [10] ? The
$f_E$ component is second class. $f_E$ has a distinctively
different reality structure, and time-reversal invariance property
versus the first class $ V, A, f_M$ [11].

{\bf (3)  Form-factor Investigations:} With respect to an
``induced" or ``effective" $t_R \rightarrow b_L$ tensorial
coupling, the physics issue becomes one of investigating and
excluding specific sources for producing such an additional
weak-moment. One of the better motivated and more developed
extensions of the SM is based on supersymmetry. The fundamental
source of SUSY breaking and its details remains an important open
problem, and the phenomenology associated with the SUSY spectra
particles is largely unknown. Nevertheless, in the present
context, SUSY provides a more general and useful off-shell
theoretical framework in which to consider these theoretical
patterns of the helicity amplitudes for $t \rightarrow W^{+} b $
decay. Form factor effects naturally occur. In particular it might
be possible to generate an additional effective $t_R \rightarrow
b_L$ weak-moment coupling of relative strength $\Lambda_{+} =
E_W/2$. For what domains, if any, of parameters in R-violating
SUSY models would this be possible? Does there exist dynamically
salient phenomena and/or persistent features besides the (i)-(v)
relations occurring in those parameter domains?  In the case of
the MSSM, a recent analysis through the one-loop level of SUSY QCD
and SUSY electroweak corrections to $\frac{\Gamma_L}{\Gamma} $,
$\frac{\Gamma_T}{\Gamma}$, and to the partial width $\Gamma$ shows
that this is not possible [12]. However, the more general case of
R-violating SUSY models remains to be investigated.

{\bf (4)  Experimental Tests/Measurements:} In on-going [1] and
forth-coming [5,6] top-quark decay experiments, important
information about the relationship of the tWb-transformation
symmetry patterns of this paper to the observed top quark decays
will come from:
\newline (a) Measurement of the sign of the $\eta_L \equiv \frac
1\Gamma |A(-1,-\frac 12)||A(0,- \frac 12)|\cos \beta _L = \pm
0.46(SM/+) $ helicity parameter [3] so as to determine the sign of
$cos \beta_L $ where $ \beta_L = \phi _{-1}^L- \phi _0^L $ is the
relative phase of the two $\lambda_b = - \frac{1}{2} $ amplitudes,
$A(\lambda _{W^{+}},\lambda _b)=|A|\exp (i\phi _{\lambda
_{W^{+}}}^{L,R})$.  \newline
 (b) Measurement of the
closely associated $ {\eta_L}^{'} \equiv \frac 1\Gamma
|A(-1,-\frac 12)||A(0,- \frac 12)|\sin \beta _L $ helicity
parameter.  This would provide useful complementary information,
since in the absence of $T_{FS}$-violation, ${\eta_L}^{'} =0$ [3].
\newline
 (c) Measurement of the partial width for $t \rightarrow
W^+ b$ such as in single-top production [13-15]. The $v^2$ factor
which differs their associated partial widths corresponds to the
SM's $\Gamma_{SM}= 1.55 GeV$, versus $\Gamma_+ = 0.66 GeV$ and a
longer-lived (+) top-quark if this mode is dominant.

{\bf Acknowledgments: }

We thank experimental and theoretical physicists for discussions.
This work was partially supported by U.S. Dept. of Energy Contract
No. DE-FG 02-86ER40291.

{\bf Appendix: The $O(v \neq y \sqrt{2}, x)$ corrections in $a$}

In this appendix is listed the form of the $O(v \neq y \sqrt{2},
x)$ corrections in $a$ as required by Lorentz invariance:

For $a=1+\varepsilon (x,y)$, the (iv) relation is $v=(1+\varepsilon )y \sqrt{2}%
m_{t}/(E_{W}+q)$ whereas from relativistic kinematics $\ \
v=q/E_{W}=[(1-y^{2}-x^{2})^{2}-4y^{2}x^{2}]^{1/2}/[1+y^{2}-x^{2}]$
. \ By equating these expressions and expanding in $x$, one
obtains $ \varepsilon =R+x^{2}S $ where
\begin{eqnarray*}
R &=&\frac{1-4y^{2}-3y^{4}-2y^{6}}{4y^{2}(1+y^{2})^{2}} \\ S
&=&\frac{-1-4y^{2}+y^{4}}{2y^{2}(1+y^{2})^{3}}
\end{eqnarray*}
and $  v= y \sqrt{2} \left[
1+R+x^{2}(S+\frac{1+R}{1-y^{2}})+O(x^{4})\right] $.  From the
latter equation, $R=(v-y\sqrt{2})/y\sqrt{2}+O(x^{2}).$

For a massless b-quark ( $x=0$ ) and $a=1$ , the (iv) relation is
equivalent
to the $\frac{m_{W}}{m_{t}}$ mass relation \  $y^{3}\sqrt{2}$ $+y^{2}+y\sqrt{%
2}-1=0$, and by relativistic kinematics to the W-boson velocity condition $%
v^{3}+v^{2}+2v-2=0$ and the simple formula $v=y\sqrt{2}$.

\newpage

\begin{center}
{\bf Table Captions}
\end{center}

Table 1:  Numerical values of the helicity amplitudes for the
standard model $(V-A)$ coupling and for the (+) coupling of
Eq.(2). The latter consists of an additional $t_R \rightarrow b_L$
weak-moment of relative strength $\Lambda_{+} \sim 53 GeV$ so as
yield a relative-sign change in the $\lambda_b= - \frac{1}{2}$
amplitudes. The values are listed first in $ g_L = g_{f_M + f_E} =
1 $ units, and second as $ A_{new} = A_{g_L = 1} / \surd \Gamma $.
Table entries are for $m_t=175GeV, \; m_W = 80.35GeV, \; m_b =
4.5GeV$.

\end{document}